\documentclass[aps,prd,groupedaddress,twocolumn]{revtex4-1}
\pdfoutput=1
\usepackage{graphicx}
\usepackage[breaklinks]{hyperref}
\usepackage[english]{babel}
\usepackage{booktabs}
\usepackage{epstopdf}
\usepackage{rotating}
\usepackage{multirow}
\usepackage{amsmath}

\begin{document}

\title{XENON1T observes tritium}

\author{Alan E. Robinson}
\affiliation{D\'epartement de physique, Universit\'e de Montr\'eal, Montr\'eal, Canada H3C 3J7}
\email{alan.robinson@umontreal.ca}

\date{June 25, 2020}

\begin{abstract}
XENON1T recently reported an excess of low-energy electron recoil events that may be attributable to either new physics or to the radioactive decay of tritium.  It is likely that hydrogen is not be effectively removed by the hot zirconium getters deployed in the detector. Cosmogenic activation of the xenon underground is found to be insufficient to describe the observed excess, although gases diffusing out of detector materials from cosmogenic activation on surface may contribute.  Changes in the operation of gas purification systems for XENON1T and other liquid nobel gas detectors could both confirm the tritium hypothesis and remove it from the detector.
\end{abstract}


\maketitle

\citet{aprile2020observation} recently reported an excess of electron recoil events below 10 keV that they attribute to either a solar axion, a neutrino magnetic moment, or tritium decays.  For the latter hypothesis, they posit that the distillation and hot getter purification systems used respectively before filling and during operation of their detector effectively remove any appreciable tritium contamination due to contamination of procured xenon or leakage of contaminated hydrogen into the detector.  Their arguments neglect tritium from fast muon spallation and the possible enrichment of tritium diffused from detector materials, although these contributions are likely negligible.  However, the system used to purify the XENON1T active volume may not be capable of removing the full amount of hydrogen that is present in XENON1T given how the purifiers are presently operated.  By adjusting the operation of the purification system, XENON1T should be able to reduce the amount of tritium in their detector.

\section{\label{sec:getter} Hydrogen purification failure modes}

The two SAES PS4-MT50-R getter modules installed on the XENON1T gas purification system \cite{XENON1T_overview} use heated sintered zirconium to irreversibly bind to oxygen and other electronegative impurities.  Hydrogen however is reversibly absorbed in hot zirconium, with hydrogen being released as the zirconium is heated \cite{Getters, GARG1990275}.  XENON1T heats their getter to 450$^\circ$C, above the specified operating temperature for these getters, in order to more efficiently crack and purify methane \cite{Getters, DOBI2010594}.  The SAES PS4-MT50-R has a separate 'Hydrogen Removal Unit' is placed inline after the hot getters, but I can not yet report any information on the performance or capacity of these units.

The purification obtainable by a hot zirconium or zirconium alloy getter is limited. The absorption of hydrogen, either that from molecular hydrogen or from water, methane, or other hydrogenated compounds, follows Sievert's law.  Using a xenon gas density of 4.54~g/L at 400$^\circ$C and 1.935~bar passing the getter, this law can be expressed as
\begin{gather}
\text{Concentration of H}_2\text{ remaining} \notag \\
= (1.0 \text{ ppb}) exp\left({\frac{5200 K}{723.15 K} - \frac{5200 K}{T}}\right) \notag \\
\times \left(\frac{\text{Concentration of H}_2\text{ absorbed}}{1\text{ ppb}} \frac{m_{Xe}}{3.2 \text{ ton}} \frac{1.14 kg}{m_{Zr}} \right )^2 
\end{gather}
for SAES St 171 getter material and in comparing to the nominal xenon mass and getter temperature in XENON1T \cite{Getters}.

The XENON1T getters would be saturated with hydrogen after absorbing an initial concentration of a few ppb of either hydrogen, methane, or water from the detector.  The hydrogen remaining in the xenon may also be preferentially enriched in tritium due to its lower diffusivity into the bulk of the zirconium getter \cite{*[{There's even a patent for using zirconium getters as an isotope separator, but the effect may be due to differences in absorption depth rather than different equilibrium sorbtion characteristics of tritium and hydrogen.}] patent}.

A significant fraction of the hydrogen in the getters may remain if they are insufficiently baked out prior to service.  For St 171 getters, heating to 900$^\circ$C while efficiently pumping out hydrogen to below 1 millitorr is recommended \cite{Getters}.  Even using this procedure, the remaining hydrogen in the getter at 450$^\circ$C produces a partial pressure of 4.0~microtorr, equivalent to 43 ppt of hydrogen in the xenon.

\section{\label{sec:tritium} Cosmogenic tritium production}

As noted by \cite{aprile2020observation} the concentration of tritium in hydrogen or water is normally at the $10^{-18}$ level.  However, this ratio is only a minimum valid for sources of tritium that are well mixed with natural sources of hydrogen.  Tritium that desorbs from materials that have been cosmogenically activated may be significantly more concentrated, thus any hydrogen off-gassed into the detector should be suspected of containing additional tritium.

While tritium and other radioisotopes can also be produced underground, this is unlikely to explain the full excess observed by XENON1T.  

It has been shown that cosmogenic activation from muon showers is largely produced by energetic secondary particles produced from the initial muon \cite{PhysRevC.71.055805}.  In light elements such as seen in Borexino \cite{Bellini_2013}, photonuclear processes dominanate, while the mass normalized production of hadrons, particularly neutrons, is enhanced for heavy target nuclei \cite{PhysRevC.96.014605}.

A full accounting of the expected tritium production in XENON1T will require a dedicated simulation of muon and muon shower spallation similar to what has been modelled for Borexino.  However, a rough estimate can be obtained by assuming that tritium is dominantly produced by neutron secondaries from muon showers.  Neutron spallation is the dominant production mechanism of cosmogenic isotopes at the earth's surface.  Following the same method to estimate the neutron spallation production cross section as in \cite{AGNESE20191} using TALYS 1.8 \cite{talys}, INCL-ABLA \cite{INCL, ABLA07}, and spectra from \citet{Gordon}, the cosmogenic production rate is estimated to be 71.5 atoms/kg/day, significantly above the values quoted by \citet{aprile2020observation} from \citet{:Zhang}. \footnote{An attempt, reported in \cite{AGNESE20191}, to reproduce the tritium production cross sections in \cite{:mei}, as cited by \cite{:Zhang}, using older versions of TALYS could not reproduce their result for neutron energies above 80~MeV.  In addition, \cite{:mei} had mistakenly overstated the exposure of the IGEX detector crystals by nearly a factor of nine \cite{:cebrian_priv}, leading to an apparent but false confirmation of their calculated results.  The apparently larger production rates from the Geant4 simulations in \cite{:Zhang} is therefore not an indication of a conservative calculation, as \cite{aprile2020observation} claims, but rather the result of a gross error in one of the compared calculations.}

\begin{figure}
\includegraphics{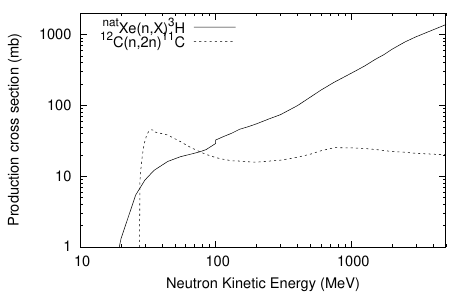}
\caption{\label{fig:1} Calculated production cross sections for tritium from neutron spallation in xenon.  TALYS 1.8 is used to calculate the cross section below 100 MeV in xenon, and below 35 MeV in carbon, while INCL-ABLA is used at higher energies.}
\end{figure}

The neutrons from muon spallation at Gran Sasso and at surface have similar spectral shapes between 100 MeV and 5 GeV, where the majority of tritium production occurs.  The neutron fluence on surfaces is however greater by a factor of $8.25\times10^6$ compared to that entering the cavern walls at Gran Sasso \cite{:mei_hime, Gordon}.  The neutron fluence in XENON1T is likely enhanced by a further factor of $(131/22.87)^{0.95}$ due to the greater neutron production efficiency in xenon versus in the rock surrounding Gran Sasso.  Combining these factors, a tritium yield of $0.036\times10^{-25}$~mol/mol$\cdot$yr results.  While this value should be considered uncertain by an order of magnitude, is is still insufficient on its own to explain the observed excess of $6.2\pm2.0\times10^{-25}$~mol/mol in XENON1T.

\section{Assay and mitigation of hydrogen and tritium}

The hot zirconium getters used in XENON1T have limited hydrogen absorption capacity and are likely to be saturated.  It should be possible to measure the amount of hydrogen in the getters by using the temperature dependence of their absorbtion capacity.  After pumping on the getters to $10^{-6}$~torr at room temperature, the getter temperature can be raised while monitoring the vacuum pressure.  A hydrogen pressure of $9.45 \times 10^{-5} \times ([\text{H}_2]/1\text{ ppb})^2$ at 450$^\circ$C can be used to determine the impurity level in the xenon.

If the contamination of the xenon is measured to be $\sim$60~ppb, as XENON1T claims would be required to explain the observed tritium signal, it would require approximately 70 kg of activated zirconium to clean their detector to less than 1~ppb of hydrogen, significantly more zirconium than is contained in the SAES PS4-MT50-R that XENON1T uses.  A purification hydrogen purification campaign, with frequent reactivation of the getters, would be required.  As the zirconium getters are easily saturated with hydrogen, any single reactivation of the getter may produce only a partial reduction of the observed tritium content.  Changing the xenon recirculation flow rate, as was done for Science Run 2 and used as a double check by XENON1T \cite{XENON1T_overview}, would have little effect on the purification of the xenon through a saturated getter.

The absorbtion efficiency of zirconium getters for hydrogen is strongly affected by the temperature in the bulk of the getter.  If a hydrogen contamination level below 1~ppb is found in XENON1T, indicating the presence of tritium enriched hydrogen, the tritium and hydrogen concentrations may be reduced by a factor of 2.9 by lowering the getter temperature from 450$^\circ$C to 400$^\circ$C.  Further reduction of the tritium content could be achieved through flushing the detector with added hydrogen.

\section{Conclusion}

The use of hot zirconium getters can remove hydrogen from nobel gases to very low concentrations, but the capacity of these systems to absorb hydrogen is limited to approximately 3~mg of hydrogen per kg of zirconium, where 3~mg is equivalent to 1 ppb of contamination in XENON1T.  Unlike for other contaminants in liquid nobel gas detectors, hydrogen is reversibly absorbed onto hot zirconium.  Therefore, the getters may be used to assay the hydrogen content of the detectors by measuring the amount of hydrogen effused during getter regeneration.  Given zirconium's low capacity for absorbing hydrogen, a campaign to purify hydrogen and tritium from XENON1T, and other liquid nobel gas detectors, is required.

\begin{acknowledgments}
Thank you to Peter Rowson and Marie-C\'ecile Piro for valuble discussions about xenon gas purification and to Susana Cebri\'an for discussions about cosmogenic activation.  This research was undertaken thanks to funding from the Canada First Research Excellence Fund through the Arthur B. McDonald Canadian Astroparticle Physics Research Institute.
\end{acknowledgments}

\bibliography{ref}

\end{document}